\documentclass[conference,10pt]{IEEEtran}
\usepackage{epsfig,rotating,setspace,latexsym,amsmath,epsf,amssymb,amsfonts,bm,subfigure,epstopdf}
\usepackage{cite,authblk}
\usepackage{bbm}
\usepackage{mathtools,amsthm}
\usepackage{algorithm}
\usepackage[noend]{algpseudocode}
\usepackage{color}
\usepackage{systeme}

\algrenewcommand\algorithmicforall{\textbf{foreach}}
\algrenewcommand\algorithmicindent{.8em}

\newtheorem{theorem}{Theorem}
\newtheorem{lemma}{Lemma}

\newenvironment{Proof}[1]{\medskip\par\noindent{\bf Proof:\,}\,#1}{{\mbox{\,$\blacksquare$}\par}}

\IEEEoverridecommandlockouts
\allowdisplaybreaks

\begin{document}

\title{Minimizing the Age of Information  Over an Erasure Channel for Random Packet Arrivals With a Storage Option at the Transmitter}

\author{Subhankar Banerjee \qquad Sennur Ulukus \qquad Anthony Ephremides\\
\normalsize Department of Electrical and Computer Engineering\\
\normalsize University of Maryland, College Park, MD 20742\\
\normalsize \emph{sbanerje@umd.edu} \qquad \emph{ulukus@umd.edu} \qquad \emph{etony@umd.edu}}
	
\maketitle

\begin{abstract}
    We consider a time slotted communication system consisting of a base station (BS) and a user. At each time slot an update packet arrives at the BS with probability $p$, and the BS successfully transmits the update packet with probability $q$ over an erasure channel. We assume that the BS has a unit size buffer where it can store an update packet upon paying a storage cost $c$. There is a trade-off between the age of information and the storage cost. We formulate this trade-off as a Markov decision process and find an optimal switching type storage policy.  
\end{abstract}

\section{Introduction}
We consider a time-slotted communication system, where at each time slot, an update packet arrives at the BS with a geometric distribution and the BS transmits the update to a user over an unreliable channel. We use the recently introduced age of information metric (see \cite{kosta2017age,YatesSurvey,SunSurvey} and the references therein) which captures the freshness of information. Most of the works in the literature which consider the minimization of age of information for such a system, i.e., a transmitter and receiver pair (can also have multiple receivers or multiple sources) over an erasure channel with stochastic arrivals of the update packet at the transmitter, has a queuing model present in it to store an update packet while it is being transmitted or to store a new update packet which may arrive while the transmitter is busy with the transmission of an older update packet, see \cite{kadota2019minimizing, saurav2022scheduling, moltafet2019closed, saurav2021minimizing, najm2018status, bedewy2019minimizing, costa2016age, saurav20223, yates2018age, moltafet2020age,najm2017status, 9478783, kaul2012status}. These papers consider a diverse set of problems, such as, deriving exact or approximated expressions for average or peak age of information for various transmission policies under different network statistics, or finding the optimal scheduling policies for various network statistics scenarios.

Similar to the aforementioned papers, we too consider that a buffer is present at the transmitter. Note that, as we are dealing with a single source and a single receiver system, and as the packet arrival time and the packet delivery time follow geometric distributions, we can show that a preemptive transmission policy is optimal; a continuous counterpart of this argument is shown in \cite{bedewy2019minimizing}. Thus, a size $1$ buffer is sufficient for our problem. However, different than all the above mentioned work, we consider that if the BS stores an update packet to the buffer it has to pay a storage cost $c$ per stored packet. 

With the addition of the storage cost, we analyze the trade-off between the storage cost and the age of information, which can be useful in the study of other problems. For example, consider a limited size buffer transmitter with multiple sources: If at time $t$, the number of packets that arrive at the transmitter is more than the available buffer size, the transmitter needs to decide which packets to store and which packets to discard. To derive a Whittle index policy \cite{whittle1988restless} for this problem necessitates a mathematical understanding of the trade-off between the storage cost and the age of information.  

In this paper, we formulate the minimization of the storage cost plus the age of information as a countable infinite state Markov decision process (MDP) and we show that there exists an optimal switching type policy for this MDP. We then find an optimal switching type storage policy. 

\section{System Model and Problem Formulation}
At each time slot, the BS receives an update packet from the process of interest with probability $0<p<1$, and the BS transmits this update packet to the user over an erasure channel with probability of successful transmission $0<q<1$. Upon the reception of the update packet, the user transmits an acknowledgement signal to the BS. We assume that this acknowledgement signal reaches the BS error-free and with a one time slot delay, i.e., if the BS transmits an update packet at time slot $t$, then at time slot $t+1$ the BS gets to know whether the update packet has been successfully received by the user or not. As at each time slot there is an uncertainty about the arrival of an update packet at the BS, and as the communication channel is erroneous, the BS may want to store the update packet. For this purpose, the BS has a unit-size buffer to store an update packet. 

We assume that each time the BS stores an update packet at the buffer it incurs a storage cost $c$. For the simplicity of this paper, we assume that the buffer drops a stored update packet at the end of one time slot after it is stored. For example, assume that at time slot $t$ the BS receives an update packet, it transmits the update packet to the user and stores the update packet in the buffer with cost $c$. Also assume that at time slot $t+1$ the BS does not receive any new update packets, however, it transmits the stored update packet from the buffer, and at the end of time slot $t+1$ the update packet gets dropped from the buffer. With these assumptions, whenever there is a successful transmission of an update packet, the age of the system drops down to either $1$ or $2$. The more general case, i.e., when a stored update packet does not get dropped from the buffer and it only gets preempted with a fresher update packet can be an interesting extension of this work; in this general case, the age may drop down to $3$, $4$, etc as well. 

Let the BS employ a storage algorithm $\pi$. We denote the action of the BS under the storage policy $\pi$ at time slot $t$ as $a^{\pi}(t)$, where $a^{\pi}(t)=1$ denotes that the BS stores the update packet at time $t$ and $a^{\pi}(t)=0$ denotes otherwise. From the above discussion it is evident that if the instantaneous age of the user at time slot $t$ is $v^{\pi}(t)$, then at time slot $t+1$ the instantaneous age is either $v^{\pi}(t)+1$ or $1$ or $2$, where the age $v^{\pi}(t)+1$ corresponds to no successful transmission of an update packet, the age $1$ corresponds to arrival of a fresh update packet at time $t$ and successful delivery of that packet by the end of time slot $t$, and finally, the age $2$ corresponds to no arrival of a fresh update packet at time $t$ and successful transmission of the stored update packet from the buffer.

A storage policy $\pi$ is completely defined by the sequence $\{a^{\pi}(t)\}_{t=1}^{\infty}$. We define the storage cost corresponding to the policy $\pi$ at time slot $t$ as $c^{\pi}(t)$, where $c^{\pi}(t) = c$ if $a^{\pi}(t) = 1$ and $c^{\pi}(t)=0$ if $a^{\pi}(t)=0$. The goal of the transmitter is to find the storage policy $\pi$ which minimizes the age of information plus the storage cost. We only consider the \emph{causal} policies, i.e., the policies which make a decision based on only the current and the past information. We define the set ${\Pi}$ as the set of all causal functions. Formally, the BS considers the following problem
\begin{align}\label{prob1}
    \inf_{\pi\in{\Pi}} \limsup_{T\rightarrow\infty} \frac{1}{T} \mathbb{E}_{\pi} \left[\sum_{t=0}^{T-1} v^\pi(t) + c^{\pi}(t)\right]
\end{align}

\section{Algorithm and Analysis}
Note that the problem (\ref{prob1}) is a countably infinite state MDP. We first define necessary components to solve this MDP.

\textit{State:} The state of the MDP in (\ref{prob1}) is $S= (v, \lambda, b)$. The component $v$ corresponds to the instantaneous age of the user. The component $\lambda$ corresponds to the availability of a fresh update packet to the BS, $\lambda = 1$ denotes the availability of a fresh update and $\lambda=0$ denotes otherwise. The component $b$ corresponds to the buffer state, $b=1$ denotes that an update packet is present in the buffer and $b=0$ denotes that there is no update packet stored in the buffer. We define $\mathcal{S}$ as the set of all possible states. Note that as the age can be unbounded, the set $\mathcal{S}$ is countably infinite. For a storage policy $\pi$, the state of the MDP at time $t$ is $S^{\pi}(t) = (v^{\pi}(t), \lambda(t), b^{\pi}(t))$. 

\textit{Transition probability:} Because of the system model, the considered problem has a time-invariant transition probabilities, i.e., the transition probabilities only depend on the state and the action, and is invariant of the time. Consider two sates in the state space $\mathcal{S}$: $S=(v,\lambda,b)$ and $S'=(v',\lambda',b')$. Under action $a$, we define the transition probability from state $S$ to state $S'$ as $P_{a}(S,S')$. First, we consider that $S = (v,1,b)$, where $b\in\{0,1\}$, $S' = (1,\lambda',b')$ and $S'' = (v+1,\lambda',b')$, where $\lambda'\in{\{0,1\}}$ and $b'\in{\{0,1\}}$, as follows
 \begin{align}
     & P_{1}\left(S,S'\right) = b' \left(q\left(\lambda' p + (1-\lambda')(1-p)\right)\right) \label{eq:3} \\ 
     & P_{0}\left(S,S'\right) = (1-b') \left(q\left(\lambda' p + (1-\lambda')(1-p)\right)\right) \label{eq:4}  \\ &  P_{1}\left(S,S''\right) = b' \left((1-q)\left(\lambda' p + (1-\lambda')(1-p)\right)\right)\label{eq:5} \\
     & P_{0}\left(S,S''\right) = (1-b') \left((1-q)\left(\lambda' p + (1-\lambda')(1-p)\right)\right) \label{eq:6} 
 \end{align}
Now, consider $S = (v,0,b)$, $S' = (2,\lambda', 0)$ and $S'' = (v+1, \lambda',0)$, where again  $\lambda'\in{\{0,1\}}$ and $b\in{\{0,1\}}$, for any action $a\in{\{0,1\}}$,
\begin{align}
\begin{split}
     P_{a} \left(S,S'\right) ={}&  b \left(q\left(\lambda' p + (1-\lambda')(1-p)\right)\right) \label{eq:7}
\end{split}\\ 
P_{a} \left(S,S''\right) ={}&(1-b) \left(\lambda' p + (1-\lambda')(1-p)\right)  \nonumber\\&
+ b \left((1-q)\left(\lambda' p + (1-\lambda')(1-p)\right)\right)\label{eq:8}
\end{align}
For any other pair of states, other than the pairs considered in (\ref{eq:3})-(\ref{eq:8}), the transition probability is $0$. 

\textit{Stationary policies:} If a policy $\pi$ is independent of time and depends only on the state of the system, then it is stationary. 

\textit{Cost:} The cost of this MDP is defined as the sum of the storage cost and the age of information. If the state of the system is $S$ and the action is $a$, where $S\in{\mathcal{S}}$ and $a\in{\{0,1\}}$, then we define the cost as $C(S,a)$. Thus,
\begin{align}\label{eq:cost}
    C(S, a) = & a c + \sum_{S'\in{\mathcal{S}}} v' P_{a}(S,S')
\end{align}
Similarly, for a policy $\pi$, the cost of the system at time slot $t$ is $C(S^{\pi}(t), a^{\pi}(t))$.

Now, consider the following optimization problem,
\begin{align}\label{prob2}
    \inf_{\pi} \limsup_{T\rightarrow\infty} \frac{1}{T} \mathbb{E}_{\pi}\left[\sum_{t=0}^{T-1} C(S^{\pi}(t),a^{\pi}(t))\right]
\end{align}
As we assume that the age of the user at time slot $1$ is $1$, it is immediate that (\ref{prob1}) and (\ref{prob2}) are equivalent.  

For an $\alpha$ such that $0<\alpha<1$, we define the total discounted cost for a policy $\pi$ with $S^{\pi}(1)=S$, where $S\in{\mathcal{S}}$, as
\begin{align}
    V_{\alpha}^{\pi}(S) = \mathbb{E}_{\pi} \left[\sum_{t=0}^{\infty} \alpha^{t} C(S^{\pi}(t),a^{\pi}(t)) \Big{|} S^{\pi}(1)=S\right]
\end{align}
We define $V_{\alpha}(S) = \inf_{\pi\in{\Pi}} V_{\alpha}^{\pi}(S)$. If there exists a policy $\pi\in{\Pi}$, for which $V_{\alpha}^{\pi}(S)$ achieves the minimum, we call it an $\alpha$-optimal policy.
\begin{theorem}\label{th:1}
There exists an optimal stationary policy $\pi$, which minimizes $\limsup_{T\rightarrow\infty}\frac{1}{T} \mathbb{E}_{\pi}\left[\sum_{t=0}^{T-1} v^\pi(t) + c^{\pi}(t)\right]$.
\end{theorem}
\begin{Proof}
Consider a stationary policy $\bar{\pi}$ such that at each time slot the BS stores an update packet with probability $1/2$. For the policy $\bar{\pi}$, all the states in $\mathcal{S}$ can be viewed as states of a Markov chain and we denote this induced Markov chain for the policy $\bar{\pi}$ as $M$. The transition probability for $M$, from state $S$ to state $S'$ is $ \frac{P_{1}(S,S')+  P_{0}(S,S')}{2}$. For this Markov chain, we define the cost for state $S$ as $ \frac{C(S,1) + C(S,0)}{2}$. Note that $M$ is irreducible as all the states are reachable from all the other states and it is aperiodic as well. 

Next, we show that $M$ is a positive recurrent Markov chain. From \cite[Thm.~1.27]{serfozo2009basics}, we know that if any one of the states of a Markov chain is positive recurrent then all of the states are positive recurrent. We show that the state $(1,1,1)$ is a positive recurrent state. We define $\tau_{S,S}$ to be the time needed to reach back to state $S$ from state $S$ for the first time. Formally, $\tau_{S,S}=\inf_{t\geq 1}\{t: S(t)=S \big{|} S(0)=S\}$. We can show that $\sum_{t=1}^{\infty} t P(\tau_{(1,1,1),(1,1,1)}=t)<\infty$, where $P$ is the conventional probability measure. Thus, $M$ is a positive recurrent Markov chain. We define $m_{S,S'}$ as the expected total cost till the system reaches state $S'$ for the first time from state $S$. Now, consider that $S$ is any arbitrary state in $\mathcal{S}$ and $S'$ is $(1,1,1)$. Note that if the BS never schedules an update packet from the buffer, i.e., if there is no fresh update packet, then the BS does not transmit, thus the average cost to go to the state $(1,1,1)$ for the first time increases. Thus, we can get an upper bound on $m_{S,(1,1,1)}$, for all ${S\in{\mathcal{S}}}$, and we can show that this upper bound is finite. Thus, $m_{S,(1,1,1)} < \infty$.

From \cite[Prop.~5]{sennott1989average}, we claim that $V_{\alpha}(S)$ is finite, $0<\alpha<1$ and $S\in{\mathcal{S}}$. Again from  \cite[Prop.~5]{sennott1989average}, we further claim that for every $S\in{\mathcal{S}}$, there exists a non-negative real number $M_{S}$, such that $V_{\alpha}(S) - V_{\alpha}((1,1,1)) \leq M_{S}$, $0<\alpha<1$. Further, for all $S\in{\mathcal{S}}$, there exists an action $a$, denoted as $a_{s}$, such that $\sum_{S'\in{\mathcal{S}}} P_{a_{S}}(S,S') M_{S'} < \infty$. As $V_{\alpha}(S)$ is finite, $S\in{\mathcal{S}}$, we can say that $0\leq V_{\alpha}(1,1,1)<\infty$. Let us define $V_{\alpha}(1,1,1)=N$, where $N\in{\mathbb{R}}$. Thus, $V_{\alpha}(S) - V_{\alpha}((1,1,1))\geq -N$, for all $S\in{\mathcal{S}}$. Thus, from \cite[Thm.~1]{sennott1989average}, we have that there exists an optimal stationary policy which minimizes $\limsup_{T\rightarrow\infty}\frac{1}{T} \mathbb{E}_{\pi}\left[\sum_{t=1}^{T} v^\pi(t) + c^{\pi}(t)\right]$.
\end{Proof}

\textit{Switch based policy:} We call a policy $\pi$ a switch based policy, if given that $\pi$ stores an update packet for state $S$ then $\pi$ stores an update packet for any other state $S'$ which satisfies $S'\geq S$, where the inequality is component-wise. 

Next, we present an important result for $V_{\alpha}(S)$ \cite[Prop.~1]{sennott1989average}.
\begin{lemma}
 For every state $S\in{\mathcal{S}}$ and $0<\alpha<1$, if $V_{\alpha}(S)$ is finite then the following relation holds true, $V_{\alpha}(S) = \min_{a}\{C(S,a) + \alpha \sum_{S'\in{\mathcal{S}}} P_{a}(S,S') V_{\alpha}(S')\}$.  
\end{lemma}

Next, we develop a policy iteration method to obtain $V_{\alpha}(S)$. We define for $n=0$, $V_{\alpha, 0}(S) = 0$, and for $n\geq 1$, $V_{\alpha,n}(S) = \min_{a} \{C(S,a) + \alpha \sum_{S'\in{\mathcal{S}}} P_{a}(S,S') V_{\alpha,n-1}(S')\}$, $S\in{\mathcal{S}}$; see \cite[Prop.~3]{sennott1989average}.
\begin{lemma}\label{lemma:3}
    For each $0<\alpha<1$, we have $V_{\alpha,n}(S) \rightarrow V_{\alpha}(S)$ for every $S$.
\end{lemma}

Recall that $S= (v, \lambda, b)$, where $v$ corresponds to the age, $\lambda$ corresponds to the availability of a fresh update packet at the BS, and $b$ corresponds to the availability of a stored update packet in the buffer.

\begin{lemma}\label{lemma:4}
For a fixed $\lambda$ and $b$, $V_{\alpha}(S)$ is an increasing function of $v$, i.e., if $S_{1} = (v_{1},\lambda,b)$ and $S_{2} = (v_{2}, \lambda,b)$ and if $v_{2}\geq v_{1}$, then $V_{\alpha}(S_{2}) \geq V_{\alpha}(S_{1})$. 
\end{lemma}

\begin{Proof}
    We first show that $V_{\alpha,n}(S_{2})\geq V_{\alpha,n}(S_{1})$, $n\in{\mathbb{N}}$. Then, the statement of this lemma directly follows from Lemma~\ref{lemma:3}. We show the monotonicity of $V_{\alpha,n}$, by induction. As $V_{\alpha,0}(S) =0$, $S\in{\mathcal{S}}$, for $n=1$, $V_{\alpha,1}(S) = \min_{a} C(S,a)$. Note that from (\ref{eq:cost}), it is immediate that $C(S_{2},a) \geq C(S_{1},a)$, $a\in{\{0,1\}}$. Thus, $V_{\alpha,1}(S_{2}) \geq V_{\alpha,1}(S_{1})$. Now assume that, 
    \begin{align}\label{eq:12}
    V_{\alpha,n-1}(S_{2})\geq V_{\alpha,n-1}(S_{1})
    \end{align}
    and as before $a\in{\{0,1\}}$,
    \begin{align}\label{eq:13}
    C(S_{2},a) \geq C(S_{1},a)
    \end{align}
    combining (\ref{eq:12}) and (\ref{eq:13}), we get,
    \begin{align}
        V_{\alpha,n}(S_{2}) =& \min_{a}\{C(S_{2},a) + \alpha \sum_{S'\in{\mathcal{S}}} P_{a}(S_{2},S') V_{\alpha,n-1}(S')\} \nonumber\\
         \geq & \min_{a}\{C(S_{1},a) + \alpha \sum_{S'\in{\mathcal{S}}} P_{a}(S_{1},S') V_{\alpha,n-1}(S')\} \nonumber\\ = & V_{\alpha,n}(S_{1})
    \end{align}
    completing the proof.
\end{Proof}

Let the state be $S$ and the action taken by the BS be $a$. Then, we define $V_{\alpha}(S;a)= C(S,a) + \alpha \sum_{S'\in{\mathcal{S}}} P_{a}(S,S') V_{\alpha}(S')$. Note that $V_{\alpha}(S) = \min_{a\in{\{0,1\}}} V_{\alpha}(S;a)$. Next, we state the optimality of a switch type policy. Note that, switching type policy is a stationary policy.

\begin{lemma}\label{lemma:5}
There exists a switching type policy which is optimal for $\limsup_{T\rightarrow\infty} \frac{1}{T} \mathbb{E}_{\pi}\left[\sum_{t=1}^{T} C(S^{\pi}(t),a^{\pi}(t))\right]$.
\end{lemma}
\begin{Proof}
    We first show that for $0<\alpha<1$, there exists a switch type policy which is $\alpha$-optimal. Consider that $\tilde{\pi}$ is an $\alpha$-optimal policy. Assume that $\tilde{\pi}$ stores an update packet at state $S=(v,1,0)$. Thus, $V_{\alpha}(S;1) - V_{\alpha}(S;0) \leq 0$, and 
    \begin{align}
        c + \alpha (1\!-\!q) (1\!-\!p) ( V_{\alpha}((v\!+\!1,0,1 )) \!-\!  V_{\alpha}((v\!+\!1,0,0))) \leq 0
    \end{align}
    Consider that $S'=(v+1,1,0)$. Now, if we can show that $V_{\alpha}((v+2,0,1)) - V_{\alpha}((v+2,0,0)) \leq V_{\alpha}((v+1,0,1)) - V_{\alpha}((v+1,0,0)) $, then it is evident that $V_{\alpha}(S';1) - V_{\alpha}(S';0)\leq 0$. Then, by induction, we can argue that if the BS chooses to store an update packet for state $S$, then the optimal choice for the BS is to store an update packet for state $S'$, where $S'=(v+x,1,0)$, where $x$ is a positive integer.

    Similar arguments can be made if the current state is $S=(v,1,1)$. If the current state is $S=(v,0,1)$ or $S=(v,0,0)$, then there is no update packet to store. Combining all these, we claim that there exists a switching type policy which is $\alpha$-optimal.
    Now, we show that $V_{\alpha}((v+2,0,1)) - V_{\alpha}((v+2,0,0)) \leq V_{\alpha}((v+1,0,1)) - V_{\alpha}((v+1,0,0))$:
    \begin{align}\label{eq:16}
        &V_{\alpha}((v+1,0,1)) - V_{\alpha}((v+1,0,0)) \nonumber\\ &=  2 q -  q(v+2) - p q V_{\alpha}((v+2,1,0)) + p q V_{\alpha}((2,1,0)) \nonumber\\& - (1-p) q V_{\alpha}((v+2,0,0)) + (1-p) q V_{\alpha}((2,0,0)) 
    \end{align}
    Similarly,
    \begin{align}\label{eq:17}
            &V_{\alpha}((v+2,0,1)) - V_{\alpha}((v+2,0,0)) \nonumber\\ &=  2 q - q (v+3) - pq V_{\alpha}((v+3,1,0)) + p q V_{\alpha}((2,1,0)) \nonumber\\& - (1-p) q V_{\alpha}((v+3,0,0)) + (1-p) q V_{\alpha}((2,0,0)) 
    \end{align}
    
From Lemma~\ref{lemma:4}, $V_{\alpha}((v+3,0,0))\geq V_{\alpha}((v+2),0,0)$ and $V_{\alpha}((v+3,1,0))\geq V_{\alpha}((v+2,1,0))$. Thus, from (\ref{eq:16}) and (\ref{eq:17}), we have $V_{\alpha}((v+2,0,1)) - V_{\alpha}((v+2,0,0)) \leq V_{\alpha}((v+1,0,1)) - V_{\alpha}((v+1,0,0))$.

Thus, for $0<\alpha<1$, there exists a switch type policy, $f_{\alpha}$, which is $\alpha$-optimal. Now, consider a sequence $\{\alpha_{n}\}_{n=1}^{\infty}\subset(0,1)$, such that $\lim_{n\rightarrow\infty} \alpha_{n}=1$, then from \cite[Lemma~1]{sennott1989average}, there exists a subsequence $\{\beta_{n}\}_{n=1}^{\infty}$ and a stationary policy $f$ such that $\lim_{n\rightarrow\infty} f_{\beta_{n}}(S) = f(S) $, $S\in{\mathcal{S}}$. Note that, $f$ is also a switch type policy. Now, from \cite[Thm.~1]{sennott1989average}, $f$ is an optimal policy for our problem. 
\end{Proof}

According to Lemma~\ref{lemma:5} there exists a switching type policy which is optimal. Let us call an optimal switching policy as $\pi_{1}$. From Lemma~\ref{lemma:5}, we know that if $\pi_{1}$ stores an update packet to the buffer for state $(v, 1, 0)$, then it stores an update packet to the buffer for any state $(v_{1},1,x)$, where $v_{1}\geq v$ and $x\in{\{0,1\}}$. Thus, we can characterize $\pi_{1}$ only based on the age of the system, i.e., if $\pi_{1}$ stores an available update packet when the age of the system is $v$, then it stores an available update packet when the age of the system is $v+x$, for $x\geq0$. 

\begin{figure}[t]
	\centerline{\includegraphics[width = 1 \columnwidth]{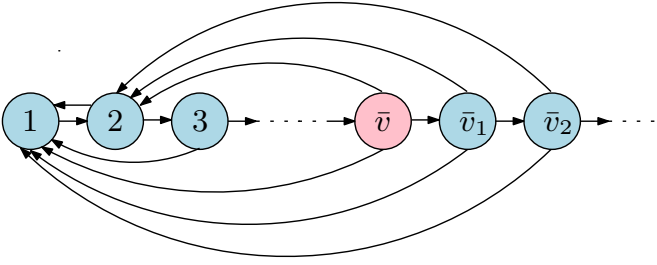}}
	\caption{Markov chain induced by $\pi_{1}$. In this figure, $\bar{v}_{1}= \bar{v}+1$ and $\bar{v}_{2}= \bar{v}+2$.}
	\label{fig1}
	\vspace*{-0.45cm}
\end{figure}

Now, consider that $\pi_{1}$ stores an available update packet when the age of the system is $\bar{v}$ or higher. If we consider a state space consisting of only the age of the system, then the policy ${\pi}_{1}$ induces an irreducible, aperiodic and positive recurrent Markov chain on that state space, see Fig.~\ref{fig1}. Thus, from \cite{serfozo2009basics}, we know that there exists a unique stationary distribution for this Markov chain. Let us denote the stationary distribution of this Markov chain as $\bm{h}$, and thus the stationary probability of the age of the system being in state $v$ as ${h}_{v}$. 

To find $\bm{h}$, we assume that the initial distribution of the Markov chain is $\bm{h}$, and as $\bm{h}$ is a stationary distribution it is a invariant distribution, i.e., if $\bm{P}$ is the probability transition matrix then $\bm{h} \bm{P}^{n} = \bm{h}$, $n\in{\mathbb{N}}$. First, we find the stationary probability of the age of the system being in state $\bar{v}+i$, $i\geq 1$,
\begin{align}\label{eq:18}
{h}_{\bar{v}+1} =  P\left(v(t)= {\bar{v}+1} | v(t-1) = \bar{v} \right) =  (1-pq) {h}_{\bar{v}}
\end{align}
For $i>1$,
\begin{align}\label{eq:19}
{h}_{\bar{v}+i} = & P(v(t) = \bar{v} + i) \nonumber\\ 
= & P(v(t) = \bar{v}+i | v(t-1) = \bar{v}+i-1) {h}_{\bar{v}+i-1 } \nonumber\\ 
= & P(v(t) = \bar{v}+i | v(t-1) = \bar{v} +i-1 , \lambda(t-1)=1) \nonumber\\ 
& \cdot P(\lambda(t-1)=1)  {h}_{\bar{v}+i-1 } + P(\lambda(t-1) = 0) {h}_{\bar{v}+i-1 } \nonumber\\ 
& \cdot P(v(t) = \bar{v}+i | v(t-1) = \bar{v} +i-1 , \lambda(t-1)=0)
\end{align}
Now, considering the individual terms in (\ref{eq:19}) we get,
\begin{align}\label{eq:20}
    P(v(t) &= \bar{v}+i | v(t-1) = \bar{v} +i-1 , \lambda(t-1)=1)\nonumber \\  
    & = (1-q)
\end{align}
and
\begin{align}\label{eq:21}
     &P(v(t) = \bar{v}+i | v(t-1) = \bar{v} +i-1 , \lambda(t-1)=0)  \nonumber\\ &= P(v(t) = \bar{v}+i | v(t-1) = \bar{v} +i-1 , \lambda(t-1)=0, \nonumber\\ & \quad \lambda(t-2) = 0) \cdot P(\lambda(t-2) = 0 | v(t-1) = \bar{v}+i-1) \nonumber\\ 
     & \quad + P(v(t) = \bar{v}+i | v(t-1) = \bar{v} +i-1 , \lambda(t-1)=0, \nonumber\\ 
     & \quad \lambda(t-2)=1) \cdot P(\lambda(t-2) = 1 | v(t-1) = \bar{v}+i-1)
\end{align}
further considering the individual terms in (\ref{eq:21}) we get,
\begin{align}\label{eq:22}
    &P(\lambda(t-2) =1 | v(t-1) = \bar{v}+i-1) \nonumber\\
    &= P(v(t-1) = \bar{v}+i-1 | \lambda(t-2)=1, v(t-2) = \nonumber\\ 
    &\quad \bar{v}+i-2) \frac{P(\lambda(t-2)=1) P(v(t-2) = \bar{v}+i-2)}{{h}_{\bar{v}+i-1}} \nonumber \\
    & = \frac{p (1-q) {h}_{\bar{v}+i-2}}{{h}_{\bar{v}+i-1}}
\end{align}
and
\begin{align}\label{eq:23}
    P(&\lambda(t-2) = 0| v(t-1) = \bar{v}+i -1) \nonumber\\ 
    & =\frac{{h}_{\bar{v}+i-1} - p (1-q) {h}_{\bar{v}+i-2} }{{h}_{\bar{v}+i-1}}
\end{align}
Substituting (\ref{eq:22}) and (\ref{eq:23}) in (\ref{eq:21}), we obtain 
\begin{align}\label{eq:24}
    P&(v(t) = \bar{v}+i | v(t-1) = \bar{v} +i-1 , \lambda(t-1)=0) \nonumber\\ 
    &= \frac{{h}_{\bar{v}+i-1} - pq(1-q){h}_{\bar{v}+i-2}}{{h}_{\bar{v}+i-1}}  
\end{align}
Finally, substituting (\ref{eq:20}) and (\ref{eq:24}) in (\ref{eq:19}), we get for $i>1$,
\begin{align}\label{eq:25}
    h_{\bar{v}+i} = & ((1-p)+ p(1-q)) h_{\bar{v}+i-1} \nonumber\\
    & - (1-p) p (1-q) q h_{\bar{v}+i -2} 
\end{align}

Therefore, ${h}_{\bar{v}+i}$ follows a recursive relation according to (\ref{eq:25}). Let $r_{1}$ and $r_{2}$ be the two roots of the following quadratic equation,
\begin{align}\label{eq:26}
    x^{2} - \left((1-p) + {p}{(1-q)}\right) x + \left((1-p) p (1-q) q\right) = 0
\end{align}
Thus, one solution to the recurrence relation in (\ref{eq:25}) is,
\begin{align}\label{eq:27}
{h}_{\bar{v}+i} = c_{1} r_{1}^{i} + c_{2} r_{2}^{i}
\end{align}
where $c_{1}$ and $c_{2}$ must satisfy the initial conditions, namely,
\begin{align}
& {h}_{\bar{v}} = c_{1} + c_{2} \label{eq:28} \\
& {h}_{\bar{v}+1} = c_{1} r_{1} + c_{2} r_{2} \label{eq:29}
\end{align}
Solving, (\ref{eq:28}) and (\ref{eq:29}) and using (\ref{eq:18}), we have
\begin{align}
   & c_{1} = \frac{(r_{2}-(1-pq)){h}_{\bar{v}}}{r_{2} -r_{1}}\\
   & c_{2} = \frac{(1-pq){h}_{\bar{v}}-r_{1}}{r_{2}-r_{1}}
\end{align}
Inserting these $c_{1}$ and $c_{2}$ into (\ref{eq:27}), we obtain, for $i\geq1$,
\begin{align}\label{eq:32}
    {h}_{\bar{v}+i} =  \left(\frac{r_{2}-(1-pq)}{r_{2}-r_{1}} r_{1}^{i} + \frac{(1-pq)-r_{1}}{r_{2}-r_{1}} r_{2}^{i} \right) {h}_{\bar{v}}
\end{align}
Note that (\ref{eq:32}) is a unique solution to (\ref{eq:25}) \cite[Thm.~2.7]{elaydi1996introduction}. We define the following quantity which we use in later calculations
\begin{align}\label{eq:imp}
 \sum_{i=0}^{\infty} {h}_{\bar{v}+i} = & {h}_{\bar{v}}\sum_{i=0}^{\infty} \left(\frac{r_{2}-1+pq}{r_{2}-r_{1}} r_{1}^{i} +\frac{1- p q - r_{1}}{r_{2}-r_{1}} r_{2}^{i}\right) \nonumber\\
 =& \frac{{h}_{\bar{v}}}{r_{2}-r_{1}} \left(\frac{r_{2}-1+pq}{1-r_{1}} + \frac{1- pq -r_{1}}{1-r_{2}}\right) \nonumber\\
 =& \frac{h_{\bar{v}}}{pq(1 + (1-p) (1-q) )} 
\end{align}

Now, when $2\leq j \leq \bar{v} $,
\begin{align}\label{eq:34}
{h}_{j} =  (1- p q)^{\bar{v}-2} {h}_{2}
\end{align}
Note that, the optimal $\bar{v}$ can never be equal to $1$, thus $\bar{v} \geq 2$, which is consistent with (\ref{eq:32}) and (\ref{eq:34}). Also,
\begin{align}
{h}_{1} = pq \sum_{i=2}^{\bar{v}-1} {h}_{i} + pq \sum_{i=0}^{\infty} {h}_{\bar{v}+i}
\end{align}
Using (\ref{eq:imp}) and (\ref{eq:34}), we get,
\begin{align}\label{eq:36}
{h}_{1} =  \left({1-(1-pq)^{\bar{v}-2}} \right) {h}_{2} + \frac{  (1-pq)^{\bar{v}-2} {h}_{2} }{1+ (1-p)(1-q)}
\end{align} 
Now, $\sum_{i=1}^{\infty} {h}_{i} =1$, and 
\begin{align}\label{eq:37}
    {h}_{2} = \frac{1} {2 - (1-pq)^{\bar{v}-2}\left(2-\frac{1}{1+(1-p)(1-q)}\left(\frac{1+ pq}{pq}\right)\right)}
\end{align}
Combining the results of (\ref{eq:32}), (\ref{eq:34}), (\ref{eq:36}) and (\ref{eq:37}), we state the following theorem.

\begin{theorem}
    For a switching type policy $\pi_{1}$ with threshold $\bar{v}$, the stationary distribution of the Markov chain induced by $\pi_{1}$ is given by,
    \begin{align}
    & {h}_{1} =  \left({1-(1-pq)^{\bar{v}-2}} \right) {h}_{2} + \frac{  (1-pq)^{\bar{v}-2}{h}_{2}}{1+ (1-p)(1-q)} \nonumber\\
    & {h}_{2} = \frac{1} {2 - (1-pq)^{\bar{v}-2}\left(2-\frac{1}{1+(1-p)(1-q)}\left(\frac{1+ pq}{pq}\right)\right)}\nonumber\\
    & {h}_{j} = (1-pq)^{\bar{v}-2} {h}_{2}, \quad 2\leq j\leq \bar{v} \nonumber \\ & {h}_{\bar{v}+i} =  \left(\frac{r_{2}-(1-pq)}{r_{2}-r_{1}} r_{1}^{i} + \frac{(1-pq)-r_{1}}{r_{2}-r_{1}} r_{2}^{i} \right){h}_{\bar{v}}, \quad i \geq 1 \nonumber
    \end{align}
\end{theorem}

Now, we find the total cost for the policy $\pi_{1}$. If the age of the system is $v$, then the corresponding cost of the system is $v+c p$, if $v\geq \bar{v}$ or else the cost is $v$. Note that, $\pi_{1}$ only stores an update packet if the system receives a fresh update packet with probability $p$. With a similar argument made for \cite[Eqn.~(3)]{malikopoulos2018average}, we can say that the cost of the $\pi_{1}$ is $\bm{h}^T\bm{c}$, where $\bm{c}$ is a vector with the $j$th component being $j$, if $j < \bar{v}$, otherwise the $j$th component is $j+c p$. If the age of the system is $\bar{v}+i$, $i\geq 0$, then the corresponding cost for $\pi_{1}$ is $\bar{v}+i+cp$. Thus, 
\begin{align}
\sum_{i=0}^{\infty} {h}_{\bar{v}+i} (\bar{v}+i+cp) = d {h}_{\bar{v}} + \frac{ {h}_{\bar{v}} \bar{v}}{p q (1+(1-p)(1-q))}
\end{align}
where $d = \frac{c}{q (1+(1-p)(1-q))} + \frac{(r_{2}-1+pq) r_{1}}{(r_{2}-r_{1})(1- r_{1})^{2}} + \frac{(1-r_{1}-pq)r_{2}}{(r_{2}-r_{1})(1- r_{2})^{2}}$.
Now, for $\bar{v}>2$ and the age of the system is $v$, $2\leq v\leq \bar{v}-1$,
\begin{align}
\sum_{v=2} ^{\bar{v}-1} {h}_{v} v =  {h}_{2} \left( \frac{pq +1}{p^2 q ^2}\right) -\frac{{h}_{\bar{v}} \bar{v}}{pq}  - {h}_{\bar{v}} \left(\frac{pq - 1 }{p^2 q^2}\right) 
\end{align}
Thus,
\begin{align}\label{eq:40}
    &\!\!\!\!\!\limsup_{T\rightarrow \infty} \frac{1}{T} \mathbb{E}_{\pi_{1}} \left[\sum_{t=1}^{T}C(S(t), a^{\pi_{1}}(t))\right] \nonumber \\ 
    =& \sum_{v=1}^{\bar{v}-1} {h}_{v} v + \sum_{i=0}^{\infty} {h}_{\bar{v}+i} (\bar{v}+i+c) \nonumber\\  
    = & d_{1} {h}_{2} + d_{2} {h}_{2}  (1-pq)^{\bar{v}-2}\bar{v} +d_{3} {h}_{2} (1-pq)^{\bar{v}-2}
\end{align}
where, $d_{1} = \frac{p^2 q^2 + p q + 1 }{p^2 q^2}$, $d_{2} = \frac{1}{pq} \left(\frac{1}{1+ (1-p)(1-q)}-1\right) $ and $d_{3}=\left( \frac{d+ d(1-p)(1-q)}{1+(1-p)(1-q)} - \frac{1 - p q + p^2 q^2}{p^2 q^2}\right)$.
Let us first see two limiting cases, i.e., $\bar{v} = 2$ and $\bar{v}=\infty$. For $\bar{v}=2$,
\begin{align}
 \limsup_{T\rightarrow \infty} &\frac{1}{T} \mathbb{E}_{\pi_{1}} \left[\sum_{t=1}^{T}C(S(t), a^{\pi_{1}}(t))\right] \nonumber \\ & = \sum_{i=0}^{\infty} {h}_{{2}+i} ({2}+i+c) + {h}_{1} \nonumber \\  & = \left(d+ \frac{1}{1+(1-p)(1-q)}\left(\frac{pq+2}{pq}\right)\right) {h}_{2}  
\end{align} 
and for $\bar{v}=\infty$,
\begin{align}
  \limsup_{T\rightarrow \infty} \frac{1}{T} \mathbb{E}_{\pi_{1}} \left[\sum_{t=1}^{T}C(S(t), a^{\pi_{1}}(t))\right] = \frac{1+pq+p^{2}q^{2}}{2 p^{2} q^{2}}
\end{align}

Now, we find extremal points of the expression of (\ref{eq:40}), i.e., we find $\bar{v}$ such that the derivative of the function $f(\bar{v}) =   d_{1} {h}_{2} + d_{2} {h}_{2}  (1-pq)^{\bar{v}-2}\bar{v} +d_{3} {h}_{2} (1-pq)^{\bar{v}-2}$ vanishes at $\bar{v}$. Replacing the expression of ${h}_{2}$ in $f(\bar{v})$, and solving it for $0$ we get the following equation of $\bar{v}$,
\begin{align}\label{eq:44}
{(1-pq)}^{\bar{v}-2} 
 =\left(\frac{d_{4}} { \left(\bar{v} + \frac{1}{\ln{(1-pq)}}-1\right)}+2\right) \frac{1}{d_{5}}
\end{align}
where, $d_{4}= \frac{\frac{p^2 q^2 +p q +1}{p^2 q^2} \left(2-\frac{pq +1}{pq(1+(1-p)(1-q))}\right) +  \frac{2}{pq(1+(1-p)(1-q))}}{\frac{1}{pq}\left(\frac{1}{1+(1-p)(1-q)}-1\right)}+2$ and $d_{5}= {2- \frac{1}{1+(1-p)(1-q)}\left(\frac{pq+1}{pq}\right)}$.

 From (\ref{eq:44}), the necessary condition for $\bar{v}$ for $f'(\bar{v}) = 0$ is,
\begin{align}
    0< &\frac{d_{4}} {d_{5}(\bar{v} + \frac{1}{\ln({1-pq)}} -1)} +  \frac{2}{d_{5}}<1
\end{align}
Now, if $\frac{d_{4}}{d_{5}}>0$, then the necessary condition for $\bar{v}$ to be an extremal point is,
\begin{align}\label{eq:46}
&\max\left\{{2,\frac{d_{4}} {d_{5}-2} - \frac{1}{\ln{(1-pq)}} +1}\right\} <\bar{v} \nonumber \\ 
&\qquad < \max\left\{2,-\frac{d_{4}} {2} - \frac{1}{\ln{(1-pq)}} +1\right\}
\end{align}
and if $\frac{d_{4}}{d_{5}}<0$, then the necessary condition for $\bar{v}$ to be an extremal point is,
\begin{align}\label{eq:47}
 &\max\left\{2,-\frac{d_{4}} {2} - \frac{1}{\ln{(1-pq)}} +1 \right\} < \bar{v} \nonumber\\ 
 &\quad < \max\left\{2,\frac{d_{4}} {d_{5}-2} - \frac{1}{\ln{(1-pq)}} +1\right\}
\end{align}
Thus, we evaluate the cost function $f(\bar{v})$, for $\bar{v}=2$, $\bar{v}=\infty$ and all the integers that satisfy (\ref{eq:46}) or (\ref{eq:47}), and choose the $\bar{v}$ which attains the minimum $f(\bar{v})$. This is the optimal $\bar{v}$ as $f$ is a continuous function. This procedure of finding optimal $\bar{v}$ is given in Algorithm~\ref{alg:1}. 

\begin{figure}[t]
\centerline{\includegraphics[width = 1\columnwidth]{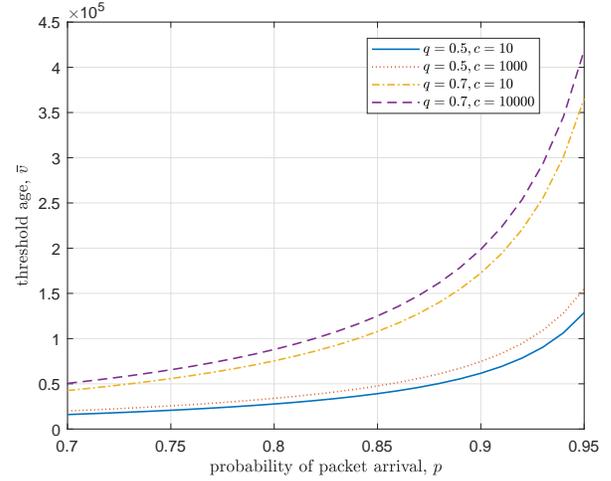}}
\vspace*{-0.3cm}
\caption{Dependence of $\bar{v}$ on varying $p$, for different $q$ and $c$.}
\label{fig2}
\vspace*{-0.5cm}
\end{figure}

In Fig.~\ref{fig2}, we plot the threshold age $\bar{v}$ as a function of $p$ for several fixed values of $q$ and $c$. Intuitively, if $p$ increases, the probability that the BS has a fresh update packet at each time slot also increases, then paying a storage cost to store an old update packet for lower values of the age becomes sub-optimal, thus the $\bar{v}$ increases with $p$. Similarly, if $q$ increases, then the probability with which a fresh update packet transmitted by the BS successfully reaches the user also increases, thus storing a stale update packet for lower values of the age becomes sub-optimal, thus the $\bar{v}$ increases with $q$. Both of these intuitions can be verified with Fig.~\ref{fig2}.

\begin{algorithm}[t]
\caption{Finding optimal $\bar{v}$}\label{alg:1}
\begin{algorithmic}
\State \textbf{Inputs}: ${h}_{2}$, $p$, $q$, $d$, $e$ 

\State \textbf{Define}: $x_{1} = \left(d+ \frac{1}{1+(1-p)(1-q)}\left(\frac{pq+2}{pq}\right)\right) {h}_{2} $ , $x_{2}=\frac{1+pq+p^{2}q^{2}}{2 p^{2} q^{2}}$, $x_{3}=\max\{2,\frac{d_{4}} {d_{5}-2} - \frac{1}{\ln{(1-pq)}} +1\}$, $x_{4}= \max\{2,-\frac{d_{4}} {2} - \frac{1}{\ln{(1-pq)}} +1 \} $,  
$x_{5}=2$,  $\mathcal{I}$ is the set of positive integers
\If{$\frac{c_{3}}{d_{5}}>0$} 
     \For{$x_{3} <\bar{v} < x_{4}$, $\bar{v} \in{\mathcal{I}}$}
        \If{$f(\bar{v})<x_{1}$}
        \State $x_{1}=f(\bar{v})$
        \State $x_{5}=\bar{v}$
        \EndIf
    \EndFor
    \Else
         \For{$x_{4} <\bar{v} < x_{3}$, $\bar{v} \in{\mathcal{I}}$}
            \If{$f(\bar{v})<x_{1}$}
            \State $x_{1}=f(\bar{v})$
            \State $x_{5}=\bar{v}$
            \EndIf
         \EndFor
   
\EndIf
\If{$x_{2}<f(x_{5})$}
    \State \textbf{Return}: $\bar{v}=\infty$
    \Else
     \State \textbf{Return}: $\bar{v}=x_{5}$
\EndIf

\end{algorithmic}
\end{algorithm}

\bibliographystyle{unsrt}
\bibliography{references}
\end{document}